\documentstyle[preprint,eqsecnum,epsf,amstex,theorem,aps]{revtex}

\newcommand{\reals}{\Bbb R}

\theoremstyle{definition}

\theoremstyle{remark}
\theoremstyle{plain}

\begin{document}
\draft
\title{Domain Wall Spacetimes and Particle Motion
}
\author{Richard Gass and Manash Mukherjee}

\address{
Department of Physics, University of Cincinnati, Cincinnati, OH
45221-0011
}

\date{\today}
\maketitle
\begin{abstract}
	
We present a mathematical framework for generating  thick domain wall solutions 
to the coupled Einstein-scalar field equations which are (locally) 
plane symmetric. This approach leads naturally to two broad classes
of wall-like solutions. The two classes include all previously known thick domain walls.
 Although one of these classes is static and the 
other dynamic, the corresponding Einstein-scalar equations share the 
same mathematical structure independent of the assumption of any 
reflection symmetry.  We also exhibit a class of thick static domain wall 
spacetimes with different asymptotic vacua.  Our analyses of particle 
motion in such spacetimes raises the interesting possibility that 
static domain walls will possess a unique experimental signature.

\end{abstract}

\narrowtext

\section{{\sf Introduction}}
Spontaneous symmetry breaking in the early universe can produce domain
walls (sometimes called cosmic membranes).  In the simplest model, a
domain wall is formed when a discrete symmetry is broken with the
resulting discrete set of vacua in different regions of space-time
\cite{1}.  The study of domain walls has a long history dating back to
the work of Zel'dovich {\it et al} \cite{2}, ( see also \cite{3} for a
recent review by Cveti\v{c} and Soleng ).  Domain walls produced at
high temperatures are cosmologically problematic since they produce
large-scale anisotropies which violate current experimental bounds
\cite{2,4,5,6}.  However, domain walls that are formed after the
decoupling of the microwave background do not produce large-scale
anisotropies {\cite 7}.  These late-time domain walls are produced at
low temperatures ( $\sim 10^{-2}$ eV) and therefore have
characteristic widths on the order of 1 MPc and low-energy densities,
thus the name soft domain walls.  Soft domain walls have large density
gradients and thus are a potential source of the density fluctuations
that are necessary for the formation of the large scale structure seen
in the universe.  It is also interesting to note that the
energy-momentum tensor for domain walls violates the strong energy
condition which results in a gravitational field that is, on average,
repulsive \cite{8,9}.

In this paper, we consider thick domain walls that arise from a single real 
scalar field with self-interaction.  The associated Einstein-Scalar 
field equations admit accelerated spherically symmetric (and hence, 
locally plane symmetric) wall-like solutions.  Unlike the previous 
investigations \cite{8,10} of other thick domain walls, we do not assume 
any reflection symmetry of the corresponding spacetime, and our 
analysis of the Einstein equations leads to two
broad classes of wall-like solutions which include previously known 
domain walls. Although non-reflection symmetric domain walls have been previously investigated by Cveti\v{c} {\it et al} and by Jensen and Soleng
\cite{3,CDGS:PRL93,CGS:PRL93,CGS:PRD93,CS:PRD95,JS:CQG97} those walls were thin domain walls. In this paper our interest is in thick domain walls.
Furthermore, we find that relaxing the assumption of reflection 
symmetry leads to a family of static plane symmetric domain walls. This result is consistent with the 
Dolgov-Khripovich theorem \cite{DK}. The Dolgov-Khripovich theorem rules out static domain walls provided the vacuum
energy density is positive and the metric has the same asymptotic limit on both sides of the wall. By violating the metric condition we are 
able to produce  explicit solutions for static domain walls. 
Linet \cite{Linet} has shown that one can also produce a static wall by using a negative vacuum energy density.
 
These walls have the unusual property that their asymptotic vacua 
differ from each other intrinsically.

In section II, we introduce a simple model of domain walls, and 
discuss the general properties of the associated energy-momentum 
tensor with particular emphasis on various energy conditions.  Without 
reflection symmetry we obtain, in section III, a new class of domain 
walls which are (locally) plane symmetric.  In section IV, the 
existence of static domain wall solutions is established.  Finally, in 
section V, we present a detailed analysis of the nontrivial particle 
motion in domain wall spacetimes.  In the Appendix, we prove two 
lemmas, which are essential for understanding the properties of static 
and accelerated domain walls.

\section{{\sf Structure of the Energy Momentum Tensor}}

Following \cite{8} we consider a real self -interacting scalar field
on a space-time $(M^{4},g)$.  The action for the scalar field is given
by
\begin{equation}
 \Lambda [\phi] = - \int_{M} \left\{ \frac{1}{2} \lambda ^{2} {\bf
 d}\phi\left(\widetilde{{\bf d}\phi}\right) + m^{4} V(\phi) \right\} \star
 {\bf 1}
\end{equation}

where $\lambda$ and $m$ set the energy scales, $\star$ is the Hodge
dual operator and the vector field $\widetilde{{\bf d}\phi}$ is the metric
dual to ${\bf d}\phi$.  [Throughout this paper, we work with units
where $\hbar = c = k_{B} = 1$ and $G=(1.2 \times 10^{19} \mbox{
GeV})^{-2}$.  In these units $\phi$ and the `potential' $V(\phi)$ are
dimensionless.] The stress energy tensor due to $\phi$ is given by
 \begin{equation}
 {\sf T} = \lambda^{2} {\bf d}\phi \otimes {\bf d}\phi -
  \left\{ \frac{1}{2} \lambda ^{2} {\bf d}\phi\left (\widetilde{{\bf
d}\phi}\right) +
  m^{4} V(\phi) \right\} g.
  \end{equation}

\noindent We assume that the potential $V(\phi)$ is positive. This ensures
that the energy density
  \begin{equation}
  {\sf T}(Z,Z) = \frac{\lambda^{2}}{2} \left\{
\left[Z(\phi)\right]^{2}+\sum_{i=1}^{3} \left[
X_{i}(\phi)\right]^{2}\right\} + m^{4} V(\phi)
  \end{equation}
  is positive. Here $\{X_0,X_1,X_2,X_3\}$ form a local orthonormal
basis and $Z=X_0$ is any observer with $g(Z,Z)=-1$.

   Since we are looking for domain wall solutions we require that
  $V(\phi)$ has two degenerate global minimum at $\phi = \mu_{\pm}$
  with $V(\mu_{\pm})=0$.  The boundary between these regions is the
  domain wall where the potential $V(\phi) > 0$ and has a maximum at
  $\phi = \mu_{0}$ .  The field $\phi$ is a smooth function on
  spacetime which
  connects the vacua $\mu_{\pm}$.  Thus, the (ideal) boundary must be
  a timelike hypersurface in $(M^{4},g)$, defined by $\phi = \mu_{0}$.
  For such field configurations the gradient field, $\widetilde{{\bf
  d}\phi}$, must be normal to the boundary hypersurface, and hence it
  is spacelike. The stress energy tensor can now be
  written as
  \begin{equation}
  {\sf T} = - \rho \left( g - \tilde{N} \otimes \tilde{N} \right)
+\nu \left( \tilde{N} \otimes \tilde{N} \right)
  \end{equation}
\noindent where the unit normal $N$ is defined by $N \equiv  \widetilde{{\bf
d}\phi}/ \left[ g \left ( \widetilde{{\bf d}\phi}, \widetilde{{\bf
d}\phi}\right) \right]^{1/2}$,
  and the functions $\rho$ and $\nu$ are given by
  \begin{equation}
  \rho =\frac{1}{2} \lambda^{2} {\bf d}  \phi \left (\widetilde{{\bf
d}\phi} \right) + m^{4} V(\phi)
  \end{equation}
  \begin{equation}
  \nu =\frac{1}{2} \lambda^{2} {\bf d}  \phi \left ( \widetilde{{\bf
d}\phi} \right) - m^{4} V(\phi).
  \end{equation}

\noindent Since $V(\phi) \geq 0$ and ${\bf d} \phi \left (
\widetilde{{\bf d}\phi} \right) >~ 0$, we have

\begin{eqnarray}
	\rho & > & 0 \\
	\rho + \nu & > & 0
\end{eqnarray}

In order to search for a domain wall configuration of the scalar field
$\phi$ as a solution to Einstein-Scalar equations $${\sf R}_{ab} -
\frac{1}{2} {\sf R} g_{ab} = (8\pi G){\sf T}_{ab}$$
we have made two physically reasonable assumptions:\\
(i) $V(\phi) \geq 0$\\
(ii) $\phi$ is spacelike, which means that $
g^{ab}\partial_{a}\phi \partial_{b}\phi > 0$.

\par \noindent These two assumptions lead to the following properties
of the stress tensor ${\sf T}$:\\
1. ${\sf T}_{ab}$ satisfies the weak-energy condition: \\
For an observer, $u^{a}$ (where $u^{a}u_{a} = -1$), we have
$$u^{a}u^{a}{\sf T}_{ab}
=\lambda^{2}(u^{a}\partial_{a}\phi )(u^{b}\partial_{b}\phi )
+\{\frac{\lambda^{2}}{2} g^{ab}\partial_{a}\phi \partial_{b}\phi
+m^{4}V(\phi)\}>0$$
2. ${\sf T}_{ab}$ satisfies the dominant-energy condition: \\
Defining $v^{a}:={\sf T}^{ab}u_{b}$, we have
$v^{a}=\lambda^{2}(\partial^{a}\phi )(\partial^{b}\phi )u_{b} -\rho
u^{a}$, which implies
$$v_{a}v^{a} = -\rho^{2} - 2m^{4}\lambda^{2} V(\phi)
(u_{a}\partial^{a}\phi )^{2}<0$$
3. ${\sf T}_{ab}$ violates the strong-energy condition,
$u^{a}u^{a}{\sf T}_{ab}\geq -\frac{1}{2}tr({\sf T})$, for
every causal vector $u^{a}$: To see this we choose an observer
$u^{a}$
with $u^{a}u_{a}=-1$ and $u^{a}\partial_{a}\phi =0$. Then, we have
$$u^{a}u^{a}{\sf T}_{ab}
=\{\frac{\lambda^{2}}{2} g^{ab}\partial_{a}\phi \partial_{b}\phi
+m^{4}{\bf V}(\phi)\}$$
$$-\frac{1}{2}tr({\sf T})
=\{\frac{\lambda^{2}}{2} g^{ab}\partial_{a}\phi \partial_{b}\phi
+2m^{4}{\bf V}(\phi)\}$$
Hence, $u^{a}u^{a}{\sf T}_{ab}< -\frac{1}{2}tr({\sf T})$,
violating the strong-energy condition for the chosen causal vector
field $u^{a}$.\\
4. Einstein's equations imply the equation of motion for the scalar
field
$\phi$:
$$0=\partial_{a}\phi {\sf T}^{ab};_{b}
=(\partial_{a}\phi \partial^{a}\phi)
\{\lambda^{2}\nabla_{b}\nabla^{b}\phi -
m^{4}\frac{d V(\phi)}{d\phi}\}$$
Here, $(\partial_{a}\phi \partial^{a}\phi)\neq 0$ since
$(\partial^{a}\phi)$ is
spacelike, and hence Einstein's equations  lead to the scalar-field
equation
$$\lambda^{2}\nabla_{b}\nabla^{b}\phi -
m^{4}\frac{d V(\phi)}{d\phi}=0$$

\section{{\sf Accelerated Spherically Symmetric Domain Walls}}

In order to determine the spacetime due to the stress-energy tensor 
(2.2) we must solve the Einstein equations
\begin{equation}
{\sf Ric} -\frac{1}{2}{\sf R} g = 8 \pi G {\sf T}
\end{equation}
where {\sf Ric} and {\sf R} are the Ricci tensor and scalar curvature 
respectively. To look for a spherically symmetric solution we choose a 
chart $(t,x,\theta,\psi)$ and 
assume that the metric
admits three Killing vector fields
\begin{eqnarray}
{\sf K_1} & = & \sin\psi \partial_{\theta} +
 \cot\theta \cos\psi \partial_{\psi}\nonumber\\
{\sf K_2} & = & -\cos\psi \partial_{\theta} +
 \cot\theta \sin\psi \partial_{\psi}\nonumber\\
{\sf K_3} & = & \partial_{\psi}
\end{eqnarray}
We consider a spherically symmetric metric of the following form:
\begin{equation}
g = f^{2}(t,x)\left\{-{\bf d}t \otimes {\bf d}t + {\bf d}x \otimes 
{\bf d}x \right\} +
 h^{2}(t,x) \left\{{\bf d}\theta  \otimes {\bf d}\theta  + \sin^{2}\theta~ {\bf 
 d}\psi  \otimes {\bf 
d}\psi  \right\}
\end{equation}
where the functions $f(t,x)$ and $h(t,x)$ are not assumed to be 
spatially reflection symmetric.
Now, taking Lie-derivatives of Einstein's equations, (3.1), with respect to 
a Killing vector field ${\sf K}$, we find
\begin{eqnarray}
	{\sf L_K T} & = & 0\nonumber\\
	{\sf L_K (trace T)} & = & 0
\end{eqnarray}
which, together with the
equations (2.4)--(2.6), gives ${\sf K (\rho)}=0$, ${\sf K (\nu)}=0$, and 
hence ${\sf K (\phi)}=0$. Then from (2.2), we have 
$\partial_\theta \phi = \partial_\psi\phi =0$. Since ${\widetilde {d\phi}}$ is spacelike,
$\phi$ is assumed to be $t$-independent in the chart $(t,x,\theta 
,\psi)$. Hence 
$d\phi=(\partial_x\phi)dx$, and the spacelike unit vector field,
$N\equiv {\widetilde {d\phi}}/[g({\widetilde {d\phi}},
{\widetilde {d\phi}})]^{\frac {1}{2}}$, normal to the membrane is given by
$N=(1/f)\partial_x$. 
Then, the membrane stress tensor (2.4) can be written as
\begin{equation}
{\sf T} = -\rho (-e^0\otimes e^0 + e^2\otimes e^2 
      + e^3\otimes e^3) + \nu e^1\otimes e^1
\end{equation}	  
where we have defined the coframes by
\begin{equation}
e^0=fdt~~;~~ e^1=fdx~~;~~ e^2=hd\theta~~;~~ e^3=h\sin \theta d\psi
\end{equation}
and the corresponding orthonormal frames are
\begin{equation}
X_0 =(1/f)\partial_t~~;
~~X_1 =(1/f)\partial_x~~;
~~X_2 =(1/h)\partial_\theta~~;
~~X_3 =(1/h\sin \theta )\partial_\psi
\end{equation}
For complete specification of the metric [(3.3)] inside a spacetime region
dominated by a thick (cosmic) membrane, it is necessary to solve for the 
functions $\{f,h\}$ from the Einstein equations
\begin{equation}
{\sf P}_a = 8\pi G \left\{{\sf T}_{ab} - ({\Gamma}/2)g_{ab}\right\}e^b
\end{equation}
where ${\sf T}$ is given by (3.5), and 
$\Gamma\equiv trace[{\sf T}]=(-3\rho + \nu)$. Now, computing the Ricci 1-forms
${\sf P}_a\equiv {\sf Ric}(X_a,X_b)e^b$ with respect to the orthonormal basis
[(3.6)-(3.7)], we find
\begin{equation}
{\sf P}_0 = Ae^0 + Be^1~~;~~
{\sf P}_1 = Be^0 + Ce^1~~;~~
{\sf P}_2 = De^2~~;~~ 
{\sf P}_3 = De^3
\end{equation}
where $\{A,B,C,D\}$ are functions of $\{t,x\}$ [Henceforth, a dot and 
a prime denote partial differentiations with respect to $t$ and $x$]. Now,
from (3.5), (3.8)-(3.9), it is seen that
$B = {\sf Ric}(X_0,X_1)=0$, and hence, in terms of the coeffecients 
$\{f,h\}$, we have 
\begin{equation}
B =-\left({\frac {2}{fh}}\right)\left\{\partial_x
\left({\frac {\dot{h}}{f}}\right) 
- \left({\frac {\dot{f}}{f^2}}\right)h'\right\}=0
\end{equation}
A simple solution to (3.10) can be given by
\begin{equation}
f\equiv f(x)~~~;~~~ h \equiv  c(t)f(x)
\end{equation}

\noindent Setting $B=0$ [(3.10)] in (3.9) and using (3.11) in the Einstein 
equations [(3.8)], we find
\begin{eqnarray}
f^2 A\equiv\partial_x\left({\frac {f'}{f}}\right) + 2\left({\frac {f'}{f}}\right)^2 
- 2\left({\frac {\ddot{c}}{c}}\right) & = & -4\pi G (\rho-\nu) f^2\nonumber\\
f^2 C\equiv -3\partial_x\left({\frac {f'}{f}}\right) & = & 4\pi G (3\rho+\nu)
f^2\nonumber\\         
-f^2 D\equiv \partial_x\left({\frac {f'}{f}}\right) + 2\left({\frac {f'}{f}}\right)^2 
- \left({\frac {\ddot{c}}{c}}\right) 
- \left({\frac {\dot{c}}{c}}\right)^2 - \left({\frac {1}{c}}\right)^2 & = & -4\pi G (\rho-\nu) f^2
\end{eqnarray}
From the first and the third equations in (3.12), we have
$({\ddot{c}}/{c}) = (1/c^{2}) +({\dot{c}}/{c})^2$ which has a solution
\begin{equation}
c(t) = {\frac {1}{k}}\cosh (kt)
\end{equation}
where $k>0$. 
Using (3.13) and defining $u=f'/f$, the  Einstein equations (3.12) reduce to 
\begin{equation}
2 u' +u^{2} -k^{2}=-(8 \pi G f^{2})\rho
\end{equation}
\begin{equation}
-3(u^{2} -k^{2}) = - (8 \pi G f^2) \nu
\end{equation}
As a consequence of (3.14)-(3.15) or $\nabla\cdot{\sf T}=0$, the equation of 
motion for the scalar field $\phi$ is given by 
\begin{equation}
\left({\frac {f'}{f}}\right) = - {\frac 13}{\frac {\nu '}{(\rho + \nu)}}
\end{equation}
It is clear from (3.5) that $\rho$ and $\nu$ represent
energy density and pressure (normal to the wall) with respect to the 
orthonormal basis
in (3.6)-(3.7) : ${\sf T}(X_0,X_0)=\rho;~~{\sf T}(X_1,X_1)=\nu$, and
by (3.14)-(3.15), $\rho\equiv\rho (x)$ and
$\nu\equiv\nu(x)$ in the chart $\{t,x,y,z\}$. [Here we remark that  
components of the stress tensor [(3.5)] are not, in general, 
time-independent with respect to the coordinate basis. For example, 
${\sf T}(\partial_y,\partial_y)=-\rho\exp(kt)$ by (3.5),(3.11) and 
(3.13)]. Furthermore,
$\nu=0$ implies $(f'/f)^2 =k^2$, and hence $\rho=0$
- leading to an empty spacetime (vacuum). Thus the pressure, ($\nu$), inside
the membranes {\sl can not be zero}.
We now show that the nowhere zero smooth function $\nu(x)<0$  based on the 
constraint $(\rho + \nu) > 0 $, and additional physical assumptions 
that $f$ is bounded and nowhere zero, and the stress tensor [(3.5)] tends to zero in the limit 
$|x|\rightarrow \infty$ giving rise to asymptotic vacua.

\par\noindent
\leftline{{\sf 3.1 Negative `Pressure' ($\nu$) :}}
\par\noindent
From equations (3.12) and (3.14)-(3.16), we have
\begin{eqnarray}
3u' & = & -4\pi G(3\rho + \nu)f^2\\
\nu' & = & -3(\rho + \nu)u
\end{eqnarray}
From the equation (3.17), it follows that $u'(x) < 0$, and hence,
$u$ has at most one
zero. Also, for the nowhere zero smooth function, $\nu$, there exists an
$\bar{x}$
 such that $\nu'(\bar{x}) =0$ (See Appendix, {\sf Lemma 2}, for a proof of
this statement). Then, (3.18) implies that each of the functions $u$
and $\nu '$ has a unique zero at $x=\bar{x}$. Furthermore,
differentiating (3.18) and using (3.17), we have
$\nu ''(\bar{x}) = -(\rho + \nu) u'(\bar{x}) > 0$ --
which means that the unique extremum of $\nu$ over $\reals$ is a
minimum at $x=\bar{x}$.
However, $\nu$ satisfies the asymptotic
condition $\lim_{|x|\rightarrow \infty}\nu=0$, and hence $\nu$ must
be {\sl negative}. As a consequence of $\nu <0$, we have $\rho > |\nu|$ since
$\rho > 0$ and $(\rho + \nu) > 0$.

\par\noindent
\leftline{{\sf 3.2 The Equation of State `$\nu \equiv \nu(\rho)$' :}}
\par\noindent

In order to solve the Einstein-scalar equations (3.14)-(3.15), we need
an equation of state $\nu \equiv \nu(\rho)$ subject to the
constraints $\nu <0$, $\rho > |\nu|$, $\lim_{|x|\rightarrow \infty}\nu
= 0~~~$ and $~~~\lim_{|x|\rightarrow \infty}\rho = 0$. First, we
observe that $u'(x) < 0~ \forall x$, and hence, from (3.15),
$\lim_{x\to \pm \infty}u(x) = \mp k$.
From these properties of $u(x)$, it
follows that the non-zero smooth functions $(u^2-k^2)$ and $u'$ are both
negative; each of them has a unique minimum and tends to zero
in the limit $|x|\rightarrow\infty$.  Using (3.14) and (3.15), we
also have
\begin{equation}
{\frac{u'(x)}{{k^2} - {{u(x)}^2}}} =  - \epsilon (x)
\end{equation}
\noindent where the nowhere zero smooth function $\epsilon (x)$ is given by
\begin{equation}
\epsilon (x)
= \frac{1}{2}\left [ \frac{3\rho (x)}{|\nu (x)|} - 1\right ] > 1
\end{equation}
since $\rho (x) > |\nu (x)|$. In fact, the equations (3.19) and (3.20) lead  to an equation of
state $\nu = - v_{0}^{2} \rho$ where
$\left(3- v_{0}^{2} \right) /2 v_{0}^{2} = \epsilon (x) > 1$, and
hence, $v_{0}^{2} <1$. It is important to note that $\epsilon (x)$ is
dimensionless according to our choice of units [specified at the
beginning of section II]. Thus, $\epsilon (x)$ is, in fact, a
function of the dimensionless argument, $kx$, where the
dimension of $k$ is given by inverse of length.

\par
To determine the classes of admissible
$\epsilon (x)$ leading to domain wall configurations of $\phi(x)$, we
recall from section II [and the discussion below (3.3)] that
$\lim_{x\rightarrow \pm\infty}\phi(x) = \mu_{\pm}$, where the constants
$\mu_{+}$ and $\mu_{-}$ represent two degenerate vacua with
$V(\mu_{\pm}) = 0$.
However, from the equations (2.5)-(2.6), (3.3) and
(3.14)-(3.15), and the existence of asymptotic limits of $\phi$, we have
\begin{eqnarray}
	\phi'(x)^{2}
	& = & (2/\lambda^{2} 8\pi G) \left [\epsilon (x) - 1\right]
	\left [k^{2} - u(x)^{2}\right ]\nonumber\\
	\lim_{|x|\rightarrow \infty}\phi' (x) & = & 0
\end{eqnarray}
\noindent where
$\lim_{|x|\rightarrow \infty}\left [k^{2} - u(x)^{2}\right ] = 0$.
Thus, any $\epsilon (x)$ from the class of ${\bf bounded}$
smooth functions such that $1~<\epsilon (x)~\leq \epsilon_{0}$,
will satisfy the condition (3.21).

\par
One can also choose $\epsilon (x)$ to be ${\bf unbounded}$. In this case, we
have another family of admissible smooth functions satisfying (3.21)
provided, asymptotically, $\left [k^{2} - u(x)^{2}\right ]$ approches
zero faster than $\epsilon (x)$ tends to infinity. To see this, we
integrate (3.19) to obtain $u(x)$:
\begin{equation}
	u(x) = -k \tanh \left (k\int \epsilon(x)~dx + c\right)\nonumber
\end{equation}
\noindent where $c$ is an integration constant. Inserting the
above expression for $u(x)$ in
$\left [k^{2} - u(x)^{2}\right ]~$, we find
\begin{equation}
	\left [k^{2} - u(x)^{2}\right ]
	= k^{2} \mbox{sech}^{2} \left (k\int \epsilon(x)~dx + c\right)\nonumber
\end{equation}
The first equation in (3.21) now gives
\begin{equation}
\phi'(x)^{2}
	 =  (2k^{2}/\lambda^{2} 8\pi G) \left [\epsilon (x) - 1\right]
\mbox{sech}^{2} \left (k\int \epsilon(x)~dx + c\right)\nonumber
\end{equation}
Since $\mbox{sech}^{2} \left (k\int \epsilon(x)~dx + c\right)$ tends to
zero exponentially as $|x|\rightarrow \infty $, we may take
$\epsilon (x)$ to be, for example, a polynomial such as $2 + 3k^{2}x^{2}$ 
for which (3.20) and (3.21) are satisfied.
\par
This suggests a procedure for
solving the Einstein equations, (3.14)-(3.15), with a choice of
$\epsilon (x)$ from these two large classes of functions leading to
domain walls. In the following subsection, we will discuss solutions
due to bounded $\epsilon (x)$.

\par\noindent
\leftline{{\sf 3.3 Solutions to the Einstein Equations :}}
\par\noindent

\noindent A simple solution to the Einstein equations, (3.14) and
(3.15), is
obtained by choosing $\epsilon (x) \equiv \epsilon_{0} > 1$ - which
trivially belongs to the class of smooth bounded functions. Hence, such
a choice ensures, through (3.21), the existence of a domain wall
solution to the Einstein equations.
Integrating (3.19) and using $u\equiv f'/f$, we find
\begin{eqnarray}
 u(x) & = & -k \tanh( \epsilon_{0} k x)\nonumber\\
 f(x) & = &\cosh^{-q}( k x/q)
\end{eqnarray}
where $q \equiv 1/ \epsilon_{0}$. Also, from (3.21) and (3.22),
we have the following exact solutions for $\phi$ and $V(\phi)$ :
\begin{eqnarray}
\phi & = & \arctan[\sinh (kx/q)]\nonumber\\
V(\phi) & = & \{\cos\phi\}^{2(1-q)}\nonumber
\end{eqnarray}
where the parameters $q$ and $k$ are related to the energy scales
$\lambda$ and $m$ by
\begin{eqnarray}
\lambda^2 & = & 2q(1-q)(1/8\pi G)\nonumber\\
m^4 & = & k^2({\frac {1}{q}} + 2)(1/8\pi G)\nonumber
\end{eqnarray}
Thus, $f(x),~\phi(x)$ and $V(\phi)$ solve the Einstein equations
as well as the field equation for the real scalar $\phi$. Finally,
the metric for the accelerated spherically symmetric domain wall
spacetime is given by
\begin{equation}
g=f^{2}(x) \left\{ - {\bf d}t \otimes {\bf d}t + {\bf d}x \otimes
{\bf d}x \right\} + (\frac{1}{k} \cosh(k t))^{2} f^{2}(x) \left\{ {\bf d}\theta \otimes
{\bf d}\theta +\sin^{2}\theta {\bf d}\psi \otimes {\bf d}\psi \right\}
\end{equation}

It is interesting to note that for a fixed $x=x_{0}$, the spherical
domain represented by (a 3-dimensional De-Sitter slice)
\begin{equation}
g|_{x=x_{0}}=f^{2}(x_{0}) \left\{ - {\bf d}t \otimes {\bf d}t +
(\frac{1}{k} \cosh(k t))^{2}({\bf d}\theta \otimes
{\bf d}\theta +\sin^{2}\theta {\bf d}\psi \otimes {\bf d}\psi )\right\}
\end{equation}
first contracts and then expands. Furthermore, if we choose a chart
$\{\tau,\eta,y,z\}$ with the following (implicit) coordinate transformations
\begin{eqnarray}
\tau  & = & (\frac{1}{k}) \ln ({\frac {1}{k}}\sinh kt + {\frac {1}{k}}\cosh kt \cos \theta) \nonumber\\
\eta & = & x \nonumber\\
y & = & (\cosh kt \sin \theta \cos \psi)/ (\sinh kt +
\cosh kt \cos \theta) \nonumber\\
z & = & (\cosh kt \sin \theta \sin \psi)/ (\sinh kt +
\cosh kt \cos \theta)
\end{eqnarray}
the wall spacetime (3.23) becomes (locally) plane symmetric:
\begin{equation}
g=f^{2}(\eta) \left\{ - {\bf d}\tau \otimes {\bf d}\tau + {\bf d}\eta  \otimes
{\bf d}\eta  \right\} + \exp (2k\tau) f^{2}(\eta ) \left\{ {\bf d}y \otimes
{\bf d}y +{\bf d}z \otimes {\bf d}z\right\}
\end{equation}

The metric (3.23) or (3.26) is only one of many possible domain wall solutions
and the equations
(3.19)-(3.20) show that admissible choices for
$\epsilon (x)$ generate additional solutions. For example, the
smooth function
$\epsilon (x) = 2\left [ 1 - kx \exp(-k^{2} x^{2})\right ]>1$ has
aymptotic limits, $\lim_{|x|\rightarrow \infty}\epsilon (x) = 2$.
Thus, $\epsilon (x)$ belongs to the class of bounded smooth functions
so that (3.21) is satisfied, and hence, $\epsilon (x)$
must lead to a domain wall solution to the Einstein equations.

The equation (3.19) for $u(x)$ then
becomes
 \begin{equation}
 {\frac{u'(x)}{{k^2} - {{u(x)}^2}}} = - 2\left [ 1 -kx
\exp\left( -k^{2} x^{2} \right) \right ].
 \end{equation}
By integrating (3.27), we get
 \begin{equation}
 u(x) = -k\tanh \left[2kx + \exp(-k^{2} x^{2}) - C\right]
 \end{equation}
 where C is a constant of integration which we will set to zero. The constant $C$ determines where $u(x)$ crosses the x-axis. By adjusting $C$ the function $u(x)$
 can be made to cross the x-axis at the origin.
Since $u=f'/f$ we can now solve for $f(x)$ which gives
\begin{equation}
 f(x)= \exp \left[ - k \int \tanh \left ( 2kx + \exp(-k^{2} x^{2})
 \right)~dx\right]
\end{equation}
 The integral in (3.29) which can not be evaluated analytically,
is easily done numerically.
   A plot of $f(x)$ is shown in Fig 1.

  The energy density $\rho (x)$ and the pressure $\nu (x)$ can be 
  computed from the Einstein equations (3.14) and (3.15) and are shown 
  in Fig 2 and Fig 3.The equation of state can no longer be found 
  analytically but it is possible to parametricaly plot the energy 
  density and the pressure as a function of position.  This is shown 
  in Fig 4.The non-zero area inclosed by the $\rho$ - $\nu$ diagram is 
  due to the lack of reflection symmetry in the stress tensor.  The 
  kink in the curve is due to the bump in the potential $V(\phi)$ 
  which is shown in figure 5.  The field $\phi$ which generates 
  $V(\phi)$ is shown in figure 6.

The equations (3.19)-(3.20) clearly allow one to
generate two broad classes of accelerated spherically symmetric domain wall
spacetimes but all of these solutions
lead to solition-like configurations for the scalar field $\phi$ and
have geodesics with the same general features. These solutions may or
may not be reflection symmetric in $x$-coordinate. In the next section,
we will show that static wall spacetimes cannot be reflection
symmetric, leading to non-trivial asymptotic vacuum structures.

\baselineskip 1truecm
\parskip .2truecm
\section{{\sf Static Plane Symmetric Domain Walls}} \label{sec1}
\par \noindent
We now look for static solutions to the Einstein-Scalar equations.
Imposing plane symmetry, the metric for a static spacetime takes the
following form \cite{11}:
\begin{equation}
g = f^2(x)\{-dt\otimes dt + dx\otimes dx\}
    + h^2(x)\{dy\otimes dy + dz\otimes dz\}
\end{equation}
Then the Einstein's equations corresponding to the domain wall stress
tensor are given by
\begin{equation}u' + 2u v = -(4\pi G) f^2 (\rho - \nu),\end{equation}
\begin{equation}u' + 2(v' + v^{2} - u v) = -(4\pi G) f^2 (3\rho +
\nu)
\end{equation}
\begin{equation}v' + 2 v^{2} = -(4\pi G) f^2 (\rho -
\nu)\end{equation}
where, we set $u\equiv (f'/f)$ and $v\equiv (h'/h)$.
Equations (4.2)-(4.4) also imply [see Appendix]
\begin{equation}(u' + 2 v') < 0\end{equation}
\begin{equation}\nu' = -(\rho + \nu) (u + 2v)\end{equation}
The above equations lead to a number of general properties of the
plane symmetric static domain wall, assuming that $(\rho + \nu) > 0$,
$\lim_{|x|\rightarrow \infty}\nu = 0$ and $\lim_{|x|\rightarrow
\infty}\rho = 0$.
\par\noindent
\leftline{{\sf 4.1 Negative `Pressure' ($\nu$) :}}
\par\noindent

From the equation (4.5), it follows that $(u + 2v)$ has at most one
zero. Also, for a nowhere zero smooth function, $\nu$, there exists an
$\bar{x}$
 such that $\nu'(\bar{x}) =0$ (See Appendix, {\sf Lemma 2}, for a proof of
this statement). Then, (4.6) implies that each of the functions $(u +
2v)$
and $\nu '$ has a unique zero at $x=\bar{x}$. Furthermore,
differentiating (4.6) and using (4.5), we have
$\nu ''(\bar{x}) = -(\rho + \nu) [u'(\bar{x}) + 2v'(\bar{x})] > 0$ --
which means that the unique  extremum of $\nu$ over $\reals$ is a
minimum at $x=\bar{x}$.
However, $\nu$ satisfies the asymptotic
condition $\lim_{|x|\rightarrow \infty}\nu=0$, and hence $\nu$ must
be {\sl negative}.
\par\noindent
\leftline{{\sf 4.2 Plane Symmetric Static Wall is Not Reflection
Symmetric :}}

\noindent
If the metric (4.1) is
reflection symmetric about $x=0$ (say) then $f(x) = f(-x)$ and
hence, $f'(x) = - f'(-x)$ and  $f'(0) =0$. Similar results hold for
$h(x)$, and $u(0) =0=v(0)$. Also,
from the Einstein's equations (4.2) and (4.3), we have
$$ (v' - u') + 2v(v-u) =0$$
Integrating once gives,
\begin{equation}[v(x) - u(x)] h^{2}(x) = K/2\end{equation} for all x
where, $K$ is a constant. For $K\neq 0$, equation (4.7)
contradicts reflection symmetry of $f(x)$ and $h(x)$, and hence a
plane symmetric static wall can not have reflection symmetry. If $K=
0$,
$ h(x) $ is proportional to $f(x)$ and with the linear changes in
coordinates we have $f(x) = h(x)$ and $u(x) = v(x)$. Then from
Einstein's equations,
$$u' + 2u^{2} = -(4\pi G) f^2 (\rho - \nu)$$
$$u' + 2(v' + v^{2} - u v) = -(4\pi G) f^2 (3\rho + \nu)$$
we find
$$0\leq 3u^{2} = (8\pi G)f^{2}\nu\leq 0$$
since, $\nu < 0$ for non-zero $\nu$. Thus, for $K=0$ we have $\nu=0$,
and
hence $\rho =0$ producing a vacuum. This conclusion also means that
a conformally flat plane symmetric static metric leads to a vacuum
spacetime under the asymptotic conditions on $\nu$ and $\rho$.
\par\noindent
\leftline{{\sf 4.3 Asymptotic Limits of The Einstein Equations :}}
\par\noindent
If we assume, again, that  pressure ($\nu$) and energy density
($\rho$) vanish
in the region far from a thick membrane, then the following equations
$$\lim_{|x|\rightarrow \infty}\nu = 0~~~;
~~~\lim_{|x|\rightarrow \infty}\rho = 0$$
generate the vacuum Einstein's equations:
\begin{equation}u' + 2u v = 0 \end{equation}
\begin{equation}u' + 2(v' + v^{2} - u v) = 0 \end{equation}
\begin{equation} v' + 2 v^{2} = 0 \end{equation}
From these equations, we find an algebraic  relation
$$(2u + v) v = 0$$
(a) If $(h'/h) \equiv v =0$, then $ (f'/f)' \equiv u' =0$. In this
case, we may take $h(x)=1$, and on integration $f(x) = \exp(px)$,
where $p$ is a constant. Inserting the following coordinate
transformations
$$\tilde{t} ={\frac {1}{p}}\exp (pt)\sinh (px);~~
\tilde{x} ={\frac {1}{p}}\exp (pt)\cosh (px);~~
\tilde{y}=y;~~\tilde{z}=z\eqno $$
in (4.1), we have the Minkowski spacetime:
$$g_{Mink} = -d\tilde{t}\otimes d\tilde{t} + d\tilde{x}\otimes
d\tilde{x }
    + d\tilde{y}\otimes d\tilde{y }+ d\tilde{z}\otimes d\tilde{z}$$
\noindent
(b) If $2(f'/f) + (h'/h) \equiv 2u + v =0$, by integrating once we
have
$f^{2}(x) h(x) = C_{0}$, where $C_{0}$ is a constant. Also,  (4.10)
implies
$h^{2}(x) = x$, and
choosing $C_{0} = 1$ we get $f^{2}(x) = 1/\sqrt{x}$. In this case, we
have Taub's plane symmetric vaccum spacetime \cite{11}:
$$g_{Taub} = {\frac {1}{\sqrt{x}}}\{-dt\otimes dt + dx\otimes dx\}
    + x\{dy\otimes dy + dz\otimes dz\}$$
\par\noindent
\leftline{{\sf 4.4 Solutions to Einstein's Equations :}}
\par\noindent
In order to obtain a plane symmetric static wall solution to Einstein's
equations (4.2), (4.3) and (4.4) we now introduce a coordinate
transformation suggested by the equation (4.7) :
\begin{equation}\frac{1}{h(x)}\frac{dh(x)}{dx} =
\frac{1}{f(x)}\frac{df(x)}{dx}
+\frac{K}{2h^{2}(x)}\end{equation}
If we define the non-singular coordinate change $\xi\equiv \xi(x)$ by
\begin{equation}d\xi = \frac{dx}{h^{2}(x)}\end{equation}
then considering $f(x)$ and $h(x)$ as functions of $\xi$, we have
\begin{equation}\frac{1}{h}\frac{dh}{d\xi} =
\frac{1}{f}\frac{df}{d\xi}
+\frac{K}{2}\end{equation}
Since $f(x)dx = f(x(\xi))	h^{2}(x(\xi))d\xi$, we define
\begin{equation}f(x(\xi))\equiv F(\xi)~~ \mbox {and}~~ N(\xi) \equiv
f(x(\xi))h^{2}(x(\xi))
\end{equation}
Thus, $h^{2}(x(\xi)) = \frac{N(\xi)}{F(\xi)}$, and hence, in the new
coordinate system $\{t,~\xi,~y,~z\}$, the plane
symmetric static metric (4.1) takes the following form:
\begin{equation}g = -F^{2}(\xi)dt\otimes dt +
N^{2}(\xi)d\xi\otimes d\xi
    + \frac{N(\xi)}{F(\xi)}\{dy\otimes dy + dz\otimes
dz\}\end{equation}
Denoting differentiation with respect to $\xi$ by a $(')$, the
equation (4.13) implies
\begin{equation}
	\left (\frac{N'}{N}\right) = 
	3\left (\frac{F'}{F}\right) + K
\end{equation}
Einstein's equations corresponding to the metric (4.15) are given by
\begin{eqnarray}
	\left (\frac{F'}{F}\right)' & = &-(4\pi G) N^2 (\rho - \nu)\\
	&   &\nonumber\\
	\frac{3}{2} \left (\frac{F'}{F}\right)^{2} 
	- \left (\frac{F'}{F}\right)\left (\frac{N'}{N}\right)
-\frac{1}{2}\left (\frac{N'}{N}\right)^{2}
+ \left (\frac{N'}{N}\right)' & = & -(4\pi G) N^2 (3\rho + \nu)\\
&   &\nonumber\\
	\frac{1}{2}\left\{\left (\frac{N'}{N}\right)' 
	- \left (\frac{F'}{F}\right)'\right \} & = & -(4\pi G) N^2 (\rho - \nu)
\end{eqnarray}
From (4.17)-(4.19) and (4.16), we have
\begin{eqnarray}
	-2\left (\frac{F'}{F}\right)' + 
	\left\{3\left (\frac{F'}{F}\right)^{2} + 
	2\left (\frac{F'}{F}\right) K + K^{2}/4\right \} & = & 
	(8\pi G) N^2 \rho(\xi)\nonumber\\
	&   &\nonumber\\
	\left \{3\left (\frac{F'}{F}\right)^{2} + 
	2\left (\frac{F'}{F}\right) K + K^{2}/4\right \} & = & 
	(8\pi G) N^2\nu(\xi) 
\end{eqnarray}
If we set $r(\xi) \equiv F'(\xi)/F(\xi) + K/3$ and 
$\alpha \equiv K/6$, then the Einstein equations (4.20) becomes
\begin{eqnarray}
2r' - 3 (r^{2} -\alpha^{2}) & = & - (8\pi G N^{2}) \rho\nonumber\\
- 3 (r^{2} -\alpha^{2}) & = & - (8\pi G N^{2}) \nu 
\end{eqnarray}
From this form of the Einstein equations, we have
\begin{equation}
{\frac{r'(\xi)}{\alpha^{2} - {{r(\xi)}^2}}} = - \sigma (\xi)
\end{equation}
where the nowhere zero smooth function $\sigma (\xi)$ is given by
\begin{equation}
\sigma (\xi) 
= \frac{3}{2}\left [ \frac{3\rho (\xi)}{|\nu (\xi)|} + 1\right ] > 3
\end{equation}
since $\rho (\xi) > |\nu (\xi)|$. 
Since $\nu < 0$, we have an equation of state in the form
$\nu(\xi) = -q(\xi)~\rho(\xi)$, where the relation
$\frac{3}{2}\left [ \frac{3\rho (\xi)}{|\nu (\xi)|} + 1\right ] 
= \sigma(\xi) > 3$ is equivalent to $0<q(\xi)<1$.
Following the procedure developed in subsection {\sf 3.2}, 
for generating 
solutions to the Einstein-scalar equations,  
we obtain the simplest static domain wall solution to (4.20)-(4.21) 
when $q$ is a constant :
\begin{equation} \ln F = -\frac{L}{3}\ln \left[ \cosh
\left(\frac{K\xi}{2L}\right) \right] -
\frac{K\xi}{3}\end{equation}
where $L$ is related to $q$ by $q = L /(2-L)$ and hence, $0<L<1$.
Now, using the equation (4.16) we find the metric functions:
\begin{equation}F^{2} = \left (\cosh (K\xi/2L)\right)^{-\frac{2L}{3}}
\exp\left (-\frac{2K\xi}{3}\right)\end{equation}
\begin{equation} N^{2} = \left (\cosh (K\xi/2L)\right)^{-2L}
\end{equation}
\begin{equation}
	\left (\frac{N}{F}\right) = \left (\cosh (K\xi/2L)\right)^{-\frac{2L}{3}}
\exp\left (\frac{K\xi}{3}\right)
\end{equation}
Equations (4.25)-(4.27) represent a two-parameter ($K,~L$) family of
plane
symmetric static spacetimes dominated by domain walls. The
`density' ($\rho$) and `pressure' ($\nu$) functions are obtained
from the Einstein's equations [4.20-4.21]:
\begin{equation}\rho(\xi) =
{{{K^2}\,\left( 2 - L \right) \,
     {{{\rm sech}\left({{K\,\xi}\over {2\,L}}\right)}^{2 -
2\,L}}}\over
   {96\,G\,L\,\pi }}\end{equation}
\begin{equation}\nu(\xi) =
{{-{K^2}\,{{{\rm sech}\left({{K\,\xi}\over {2\,L}}\right)}^{2 -
2\,L}} }
    \over {96\,G\,\pi }}\end{equation}
 The explicit forms
of the (dimensionless) scalar field and the potential are given by
\begin{equation}
\phi(\xi) =
\mbox { arctan}\left({\rm sinh}\left({{K\,\xi}\over
{2\,L}}\right)\right)\end{equation}
\begin{equation}V(\phi) = {\left(\cos \phi \right)}^{2 -
2\,L}\end{equation}
The energy scales, $\lambda$ and $m$, are related to the parameters,
$K$ and $L$, by 
\begin{equation}\lambda^{2} =
{{\left( 1 - L \right) \,L}\over {12\,G\,\pi }} \end{equation}
\begin{equation}m^{4} =
{{{K^2}}\over {48\,G\,L\,\pi }}
\end{equation}

\noindent It should be noted from (4.28)-(4.29) that the matter density and
pressure are reflection symmetric, while the spacetime [determined by
(4.25)-(4.27)] is not. The absence of reflection symmetry of the static
wall spacetime is also demonstrated by the existence of different
asymptotic vacua defined by
$$\lim_{|\xi|\rightarrow \infty}\nu = 0~~~;
~~~\lim_{|\xi|\rightarrow \infty}\rho = 0$$
For $\xi\rightarrow \infty$,
\begin{eqnarray}
	F^{2} & \rightarrow & \exp {(-K\xi)}\\
	N^{2} & \rightarrow & \exp {(-K\xi)}\\
	\left (\frac{N}{F}\right)    & \rightarrow & \mbox{constant}
\end{eqnarray}
Comparing our results in subection ${\bf 4.3}$, it is clear that the
limits (4.34)-(4.36) implies a Minkowski vacuum.  For $\xi\rightarrow
-\infty$,
\begin{eqnarray}
	F^{2} & \rightarrow & \exp \left (-{\frac{1}{3}}K\xi\right)\\
	N^{2} & \rightarrow & \exp {(K\xi)}\\
	\left (\frac{N}{F}\right)     & \rightarrow & 
	\exp \left ({\frac{2}{3}}K\xi\right)
\end{eqnarray}
The fact that the limits (4.37)-(4.39) yields the Taub vacuum, can be
seen as follow:
\begin{equation}
	g_{\xi\rightarrow -\infty}
	= -\exp {(-{\frac{1}{3}}K\xi)}dt\otimes dt + \exp
	{(K\xi)}d\xi\otimes d\xi + \exp
	{({\frac{2}{3}}K\xi)}\{dy\otimes dy + dz\otimes dz\}
\end{equation}
Then, the coordinate transformations given by
\begin{eqnarray}
	t & = & (2K/3)^{\frac {1}{3}} t_{o}\\
	\xi & = & (3/2K) \ln \{(2K/3)^{\frac {4}{3}} \xi_{o}\}\\
	y & = & (3/2K)^{\frac {2}{3}}y_{o}\\
	z & = & (3/2K)^{\frac {2}{3}}z_{o}
\end{eqnarray}
lead to the metric for the Taub vacuum:
\begin{equation}
	g_{Taub} = {\frac {1}{\sqrt{\xi_{o}}}}\{-dt_{o}\otimes dt_{o} +
	d\xi_{o}\otimes d\xi_{o}\} + \xi_{o}\{dy_{o}\otimes dy_{o} +
	dz_{o}\otimes dz_{o}\}
\end{equation}

\leftline{{\sf 4.5 More  Static Solutions :}}
\par\noindent In (4.22), if we take a nowhere zero bounded 
smooth function 
\begin{equation}
 \sigma(\xi) =9\left [1 + \exp(-36\alpha^{2} \xi^2)\right ]\nonumber
\end{equation}
\noindent which satisfies (4.23), then we have
\begin{equation}
  r(\xi)=-\alpha \tanh \left 
  [(9\alpha)\xi + 
  (\frac{3}{4}) \sqrt{\pi}~\mbox{Erf} (6\alpha\xi) -C\right]
\end{equation}
A plot of $r(\xi)$ is shown, for $C=0$ and $\alpha = 1/6$ 
[or equivalently, $K=1$], in figure 7.  
The function $ 
F(\xi)$ can be found by numerical integration.  A plot of $F(\xi)$ 
is shown in figure 8.  Equation 4.16 can be integrated to give 
$N(\xi)$.  A plot of $N(\xi)$ is shown in figure 9.

\section{{\sf Particle Motion}}
For domain wall spacetimes gravity is, on average, repulsive.  This
can be seen from the following argument.  The condition, ${\sf
Ric}(v,v)\geq 0$ for all observers $v$, expresses the emperical fact
that, on average, gravity attracts \cite{12}.  This condition is
equivalent to the strong energy condition, ${\sf T}(v,v)\geq
-\frac{\Gamma}{2}$ for all observers $v$, where ${\Gamma} =
\mbox{trace} ({\sf T})$.  Thus, violation of the strong energy
condition for the Einstein-Scalar stress tensor implies that, on
average, gravity is repulsive in the domain wall spacetime.

\subsection{{\sf Accelerated Spacetimes}}

To obtain the geodesic eqations for a particle (of mass $m$) in the 
accelerated domain wall  spacetime  [equation (3.23)] , we consider its worldline in the equatorial plane [$\theta = 
\pi/2$] :
\begin{equation}\gamma : \tau \mapsto (t=t(\tau),x= 
(\tau),\theta=\pi/2,\psi=\psi (\tau))
\end{equation}
If $\gamma$ is a geodesic (and $\tau$ is an affine parameter along $\gamma$), then
the 4-velocity of the particle is given by
\begin{equation}
\gamma_* = \left({\frac {dt}{d\tau}}\right)\partial_t 
+ \left({\frac {dx}{d\tau}}\right)\partial_x 
+ \left({\frac {d\psi}{d\tau}}\right)\partial_\psi 
\end{equation}
with the normalisation  $g(\gamma_*,\gamma_*)= -m^2$:
\begin{equation}
-f^{2}\left ({\frac {dt}{d\tau}}\right)^2 
+f^{2} \left({\frac {dx}{d\tau}}\right)^2 + 
 f^{2}(\cosh kt /k)^{2}\left({\frac {d\psi}{d\tau}}\right)^2= -m^2	
\end{equation}
Then the geodesic $\gamma$ is specified by
\begin{eqnarray}
\left ({\frac {dx}{d\tau}}\right)^2 & = & \left ({\frac {1}{f^{2}}}\right)
\left ({\frac {a_{0}}{f^{2}}} - m^{2}\right)\\
\left ({\frac {dt}{d\tau}}\right)^2 & = & \left ({\frac {1}{f^{2}}}\right)
\left\{a_{0} - \left ({\frac {a_{3}k}{\cosh kt}}\right)^{2}\right\}\\
\frac {d\psi}{d\tau} & = &  \left ({\frac {1}{f^{2}}}\right)
\left ({\frac {a_{3}k^{2}}{\cosh^{2}kt}}\right)
\end{eqnarray}
where $a_{0}$ and $a_{3}$ are constants of integration. For a massive 
particle ($m=1$), 3-velocity with respect to an observer 
$Z=(1/f)\partial_{t} $ is given by
\begin{equation}
	v^{2} = 1 - \left \{a_{0}/f^{2} + (a_{3}k/f\cosh kt)^{2}\right\}^{-1}
\end{equation}
 For any given domain wall solution these equations can be solved 
 numerically.  We will work with $f(x) = \cosh^{-q}( k x/q)$ [(3.22)] solution 
 but the essential features of particle motion remain unchanged in 
 other spacetimes obtained from (3.19)-(3.21).
 Since $f(x)$ decreases monotonically away from $x=0$ (where it 
 attains its unique maximum), the 3-velocity of the partcle with respect 
 to the observer $Z$, will increase away from $x=0$. This shows 
the repulsive nature of the domain wall gravitational field 
 \par Since domain wall spacetimes allow repulsive gravity it is clear 
 that ``turning point" solutions in which a incoming particle is 
 repelled by the domain wall are allowed.  The phase space plot for 
 one such geodesic is shown in figure 10.  Since the spacetime is 
 expanding and particles are accelerated away from the wall the 
 coordinate time $t$ along a geodesic is red shifted.  A plot of 
 $t(\tau)$ is shown in figure 11.

  \subsection{{\sf Static Spacetimes}}

 To analyse particle motion in a static domain wall spacetime we now 
 consider the geodesic equations for a test particle (of mass $m=0,~1$) .  
 Such a particle may be represented by a curve on 
 $(M^4,g)$ [with $g$ given by  (4.15) and (4.25)-(4.27)] :
 \begin{equation}\gamma : \tau \mapsto (t=t(\tau),\xi=\xi (\tau),y=y(\tau),z=z(\tau))
\end{equation}
If $\gamma$ is a geodesic (and $\tau$ is an affine parameter along $\gamma$), then
the 4-velocity of the particle is
 \begin{equation}
	 \gamma_* = \left({\frac {dt}{d\tau}}\right)\partial_t 
+ \left({\frac {d\xi}{d\tau}}\right)\partial_\xi + \left({\frac 
{dy}{d\tau}}\right)\partial_y  + \left({\frac {dz}{d\tau}}\right)\partial_z 
\end{equation}
with the normalisation  $g(\gamma_*,\gamma_*)= -m^2$:
\begin{equation}
-F^2\left ({\frac {dt}{d\tau}}\right)^2 
   +N^{2} \left({\frac {d\xi}{d\tau}}\right)^2 + 
   (N/F)\left\{\left({\frac {dy}{d\tau}}\right)^2 
   + \left({\frac {dz}{d\tau}}\right)^2\right\}= -m^2	
\end{equation}
The geodesic equations are given by
\begin{eqnarray}
\frac {dt}{d\tau} & = & C_{0}/F^{2}\\
\frac {d}{d\tau}\left(N^{2}{\frac {d\xi}{d\tau}}\right)  & = & 
{\frac {1}{2}} \left \{-(F^{2})'\left(\frac 
{dt}{d\tau}\right)^{2} + (N^{2})'\left (\frac 
{d\xi}{d\tau}\right)^{2} + (N/F)'\left (\frac 
{dy}{d\tau}\right)^{2} +(N/F)'\left (\frac 
{dy}{d\tau}\right)^{2}\right\}\\
\frac {dy}{d\tau} & = & C_{2}/(N/F)\\
\frac {dz}{d\tau} & = & C_{3}/(N/F)
\end{eqnarray}
where $C_{0}$, $C_{2}$, and $C_{3}$ are constant and `prime' 
represents differentiation with respect to $\xi$. 

\noindent  For a massive 
particle ($m=1$), the 3-velocity with respect to an observer 
$Z=(1/F)\partial_{t} $ is given by
\begin{equation}
	v^{2} = 1 - (F/C_{0})^{2}
\end{equation}
 Since $F$ is a monotonically decreasing positive function, a massive 
 particle that starts on the left side with a negative velocity will 
 move away from the wall, reach a turning point and be accelerated 
 back towards the wall.  After the particle has passed through the 
 wall it will be repelled from the wall.  Thus, on the left side of 
 the domain wall (Taub-vacuum), gravity is attractive while, on the 
 right side (Minkowsky-vacuum), gravity is repulsive.  A phase space 
 plot for such a particle is shown in figure 12.
 
 For a photon traveling along $\xi$-axis, the corresponding null 
 geodesic is represented by
 \begin{equation}\gamma : \lambda \mapsto (t=t(\lambda),~\xi=\xi 
 (\lambda),~y=a,~z=b)
\end{equation}
where, $\lambda$ ia an affine parameter, and $a,~b$ are constants. 
This null geodesic is characterised by 
\begin{eqnarray}
\gamma_* & = & \left({\frac {dt}{d\lambda}}\right)\partial_t 
+ \left({\frac {d\xi}{d\lambda}}\right)\partial_\xi \\	
\frac {dt}{d\lambda} & = & C_{0}/F^{2} \\
g(\gamma_*,\gamma_*) & = & 0 = -F^2\left ({\frac {dt}{d\lambda}}\right)^2 
   +N^{2} \left({\frac {d\xi}{d\lambda}}\right)^2
\end{eqnarray} 
Now, with respect to the observer field $Z=(1/F)\partial_{t}$, we 
compute the frequency ratio for the photons traveling along 
$\xi$-axis:
\begin{equation}
	\frac {f_{e}}{f_{o}} = \{- g(\gamma_{*},~Z_{e})\}/\{- 
	g(\gamma_{*},~Z_{o})\} = F(\xi_{o})/F(\xi_{e})
\end{equation}
where $e$ and $o$ are emission-point and observation-point respectively. 
From the above equation, we note that 		
photons approaching the wall from the left-hand (Taub) side 
 will be continuously blue-shifted as they pass through the wall 
 toward the Minkowski vacuum.  Photons that approach the wall from the 
 right-hand (Minkowski) side of the wall will first be red-shifted and 
 then blue-shifted when they pass through the wall into the Taub side.

The geodesic equations with respect to the 
 generalised static wall metric discussed in section {\bf 4.5}, can 
 be obtained by replacing $F$ in the above equations.
The essential characteristics of particle motion are the same as 
discussed above.      

\section{{\sf Conclusions}}

We have found new classes of static and non-static domain wall solutions 
to the coupled Einstein-scalar field equations which are (locally) 
plane symmetric.  The static walls are particularly interesting since static domains walls cannot be reflection symmetric.
Consequently, static domain walls must possess different asymptotic vacua. In principle, it should be possible to detect static
domain walls by studing the red-shift of photons. The fact that photons approaching the wall from the Taub side
 are continuously blue-shifted as they pass through the wall
 toward the Minkowski vacuum, while photons that approach the wall from the
 Minkowski side of the wall are first  red-shifted and
 then blue-shifted when they pass through the wall into the Taub side provides a unique experimental signature.
 
 It is generally believed that small perturbations in the mass 
distribution in the early Universe could subsequently grow \cite{13,14} to form 
galaxies and  the structures we currently see around us. However, one of the 
major problems \cite{15} in modern cosmology is to identify the origin of these 
perturbations. The domain walls discussed in this paper could be 
possible sources of the density fluctuations through their specific 
gravitational interactions with the ambient matter.

 \par\noindent \centerline{{\sf Acknowledgement}} \par\noindent
 MM would like to thank Mark Jarrell and Frank Pinski for 
 their help where it counts, 
 Louis Witten, Paul Esposito and Rohana Wijewardhana for their 
 interest in this work, and Robin Endelman for  
 many helpful discussions. RG would like to thank Claudia Taylor for helpful discussions.
 \newpage 
 \par\noindent \centerline{{\sf APPENDIX}} \par\noindent {\bf Lemma
 1}: $(u' + 2 v') < 0$ [Equation (4.5)]. \par\noindent
 \underline{Proof:} Equating (4.2) and (4.4) leads to
$$ 2(v^{2} - u v) = u' - v'$$
Inserting the above relation in (4.3) gives
$$u' + v'/2 = -(2\pi G)f^{2}(3\rho + \nu) < 0$$
This equation, together with $v' < 0$ [which follows from (4.4)],
proves the lemma.
\par\noindent
{\bf Lemma 2}: Let $\nu(x)$ be a smooth nowhere-zero function on ${\Bbb
R}$. If
$\lim_{|x|\rightarrow \infty}\nu=0$, then there is an $\bar{x}$ such
that $\nu'(\bar{x}) = 0$.
\par\noindent
\underline{Proof:} Suppose that $\nu' \neq 0$ for all $x\in
{\reals}$. Since
$\nu(x)$ is  smooth, $\nu'(x)$ is continuous and hence, either
$\nu'(x)>0$ or $\nu'(x)<0$.
If $\nu'(x)>0$, then $\nu$ is a strictly increasing function on
${\reals}$, and hence it can not satisfy the boundary condition
$\lim_{|x|\rightarrow \infty}\nu=0$. For $\nu'(x)<0$, a similar
argument leads to the same contradiction. Thus, there exists an
$\bar{x}\in {\reals}$ such that $\nu'(\bar{x}) = 0$.

\begin{figure}
 \epsfxsize= 3in
 \centerline{ \epsffile{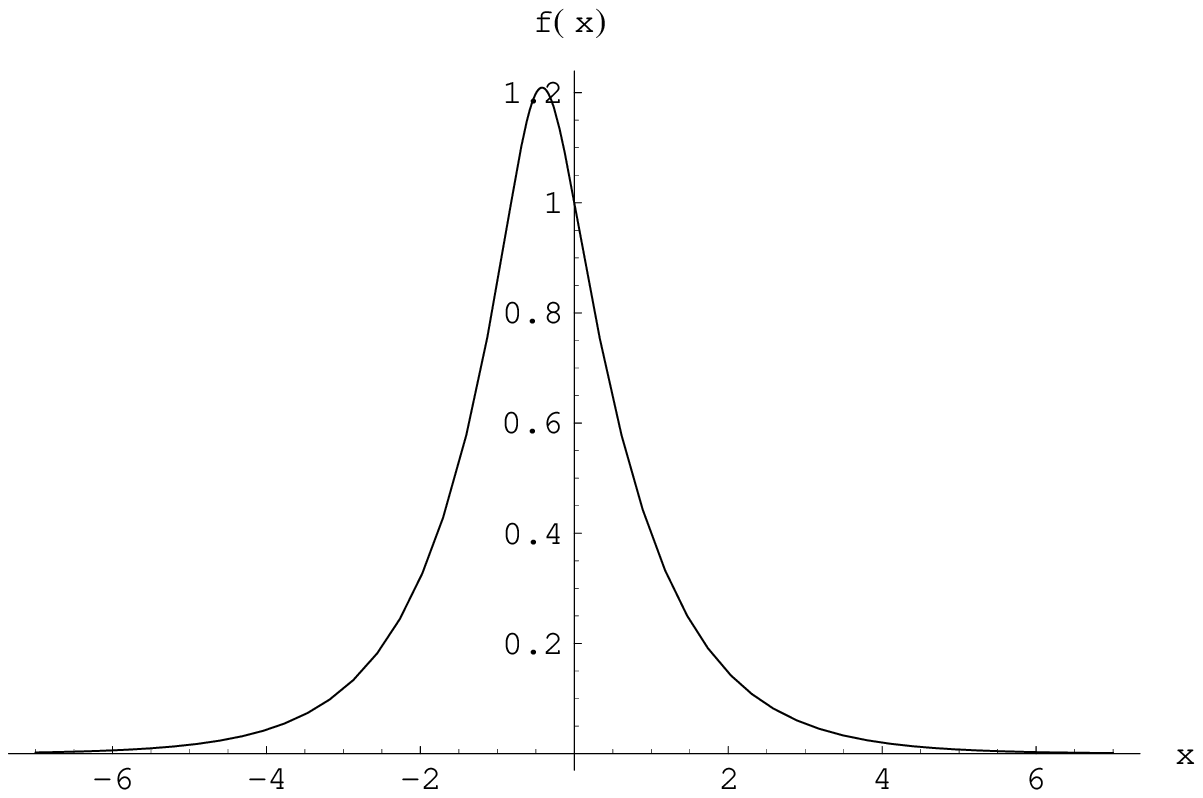}}
  
  \centerline{\caption{Plot of $f(x)$ as defined by Eq. 3.29 versus $x$.  Note that the domain wall is no longer 
  centered at $x=0$.}}

  \end{figure}
  
  \begin{figure}
 \epsfxsize= 3in
 \centerline{ \epsffile{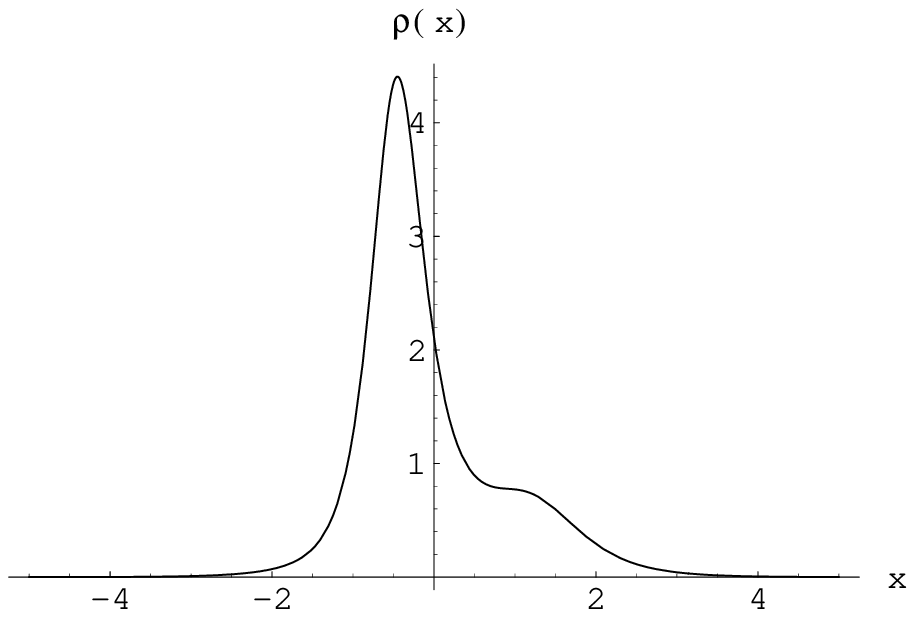}}
  
  \centerline{\caption{Plot of the energy density $\rho(x)$ versus $x$ for $k=1$.}}  
  \end{figure}
  
  \begin{figure}
 \epsfxsize= 3in
 \centerline{ \epsffile{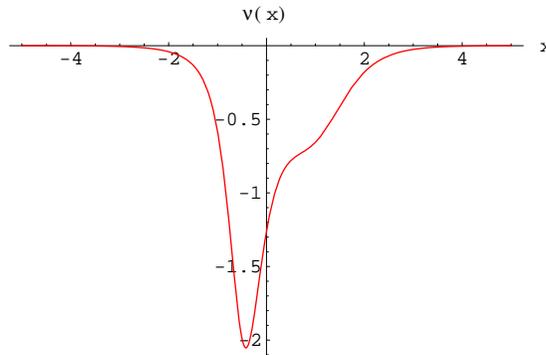}}
  
  \centerline{\caption{Plot of the pressure $\nu(x)$ versus $x$ for $k=1 $.
   Note that the pressure is negative, as it must be for a domain wall solution.}}
\end{figure}

\begin{figure}
 \epsfxsize= 3in
 \centerline{ \epsffile{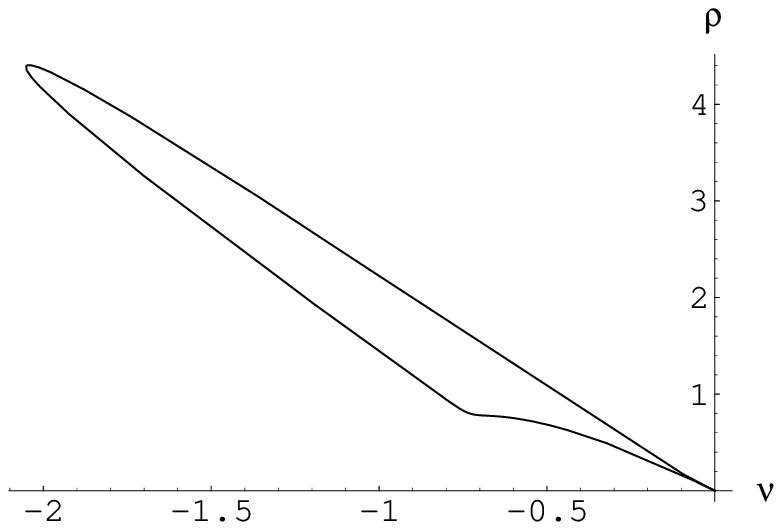}}
  
  \centerline{\caption{Plot of the energy density $\rho(x)$ versus  the pressure $\nu(x)$ for $k=1$ .}}  
  \end{figure}
  
   \begin{figure}
 \epsfxsize= 3in
 \centerline{ \epsffile{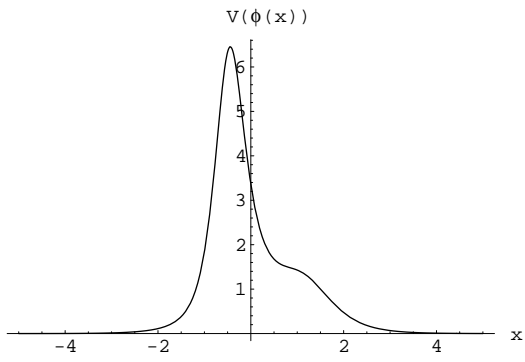}}
  
  \centerline{\caption{Plot of the potential $V(\phi)$ versus $x$ for $k=1$.}}  
  \end{figure}

 \begin{figure}
 \epsfxsize= 3in
 \centerline{ \epsffile{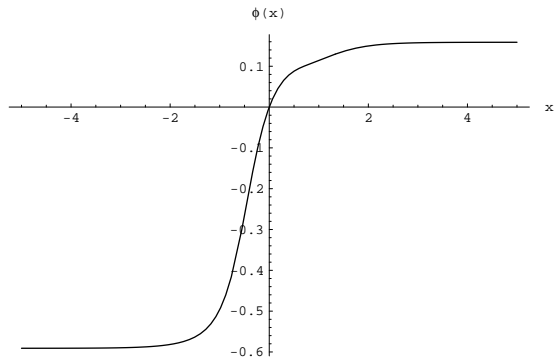}}
  
  \centerline{\caption{Plot of the field $\phi(x)$ versus $x$ for $k=1$.}}  
  \end{figure}
  
  \begin{figure}
 \epsfxsize= 3in
 \centerline{ \epsffile{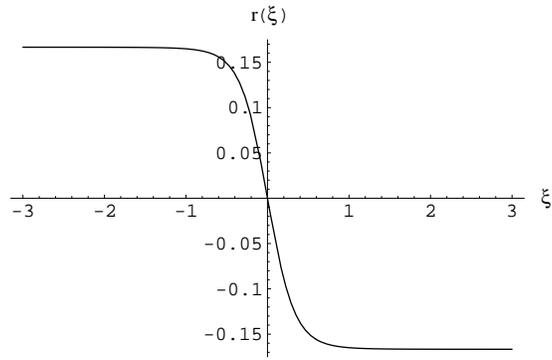}}
  
  \centerline{\caption{Plot of  $r(\xi)$  as defined by Eq. 4.46 versus $x$ for $K=1 $.}}  
  \end{figure}

  \begin{figure}
 \epsfxsize= 3in
 \centerline{ \epsffile{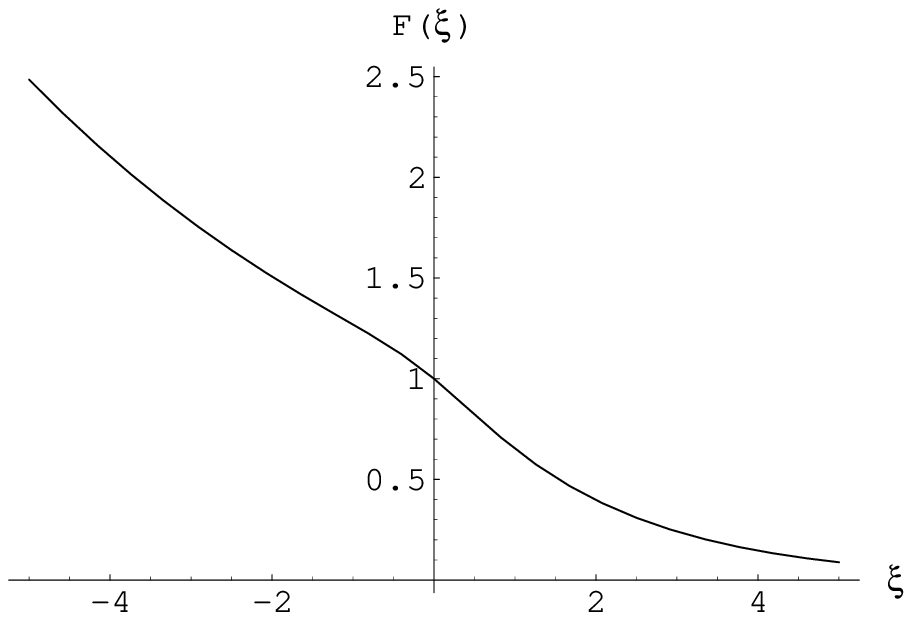}}
  
  \centerline{\caption{Plot of  $F(\xi)$ versus $\xi$ for $K=1$.}}  
  \end{figure}
  
   \begin{figure}
 \epsfxsize= 3in
 \centerline{ \epsffile{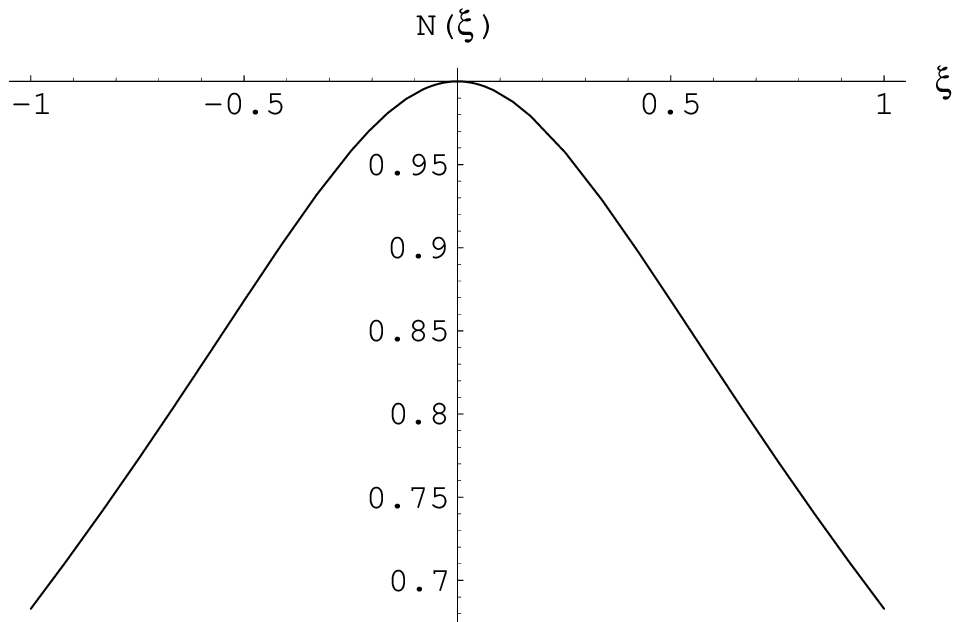}}
  
  \centerline{\caption{Plot of  $N(\xi)$ versus $\xi$ for $K=1$. }}  
  \end{figure}
  
   \begin{figure}
 \epsfxsize= 3in
 \centerline{ \epsffile{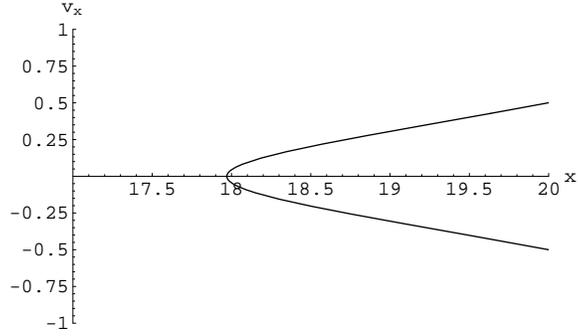}}
  
  \centerline{\caption{Phase space plot of a turning point solution for a  particle with initial conditions
   $x(0)=20,x^{'}(0)=-0.5,y^{'}(0)=0.5,y(0)=0,z^{'}(0)=0,
  z(0)=0,t^{'}(0)=1,t(0)=0$ and f given by 3.22.}}  
  \end{figure}
  
  
  
  
  
  
  
   \begin{figure}
 \epsfxsize= 3in
 \centerline{ \epsffile{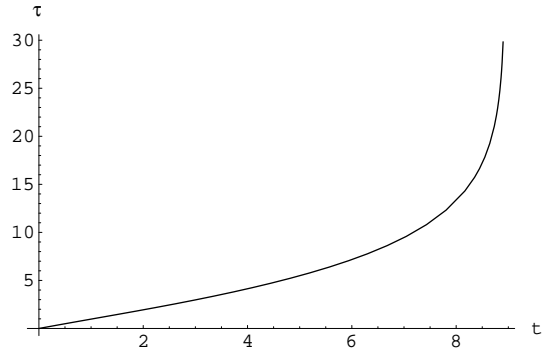}}
  
  \centerline{\caption{A plot of $\tau(t)$ showing the redshift of the t for a particle with initial 
  conditions $x(0)=1/2,x^{'}(0)=-0.5,y^{'}(0)=0,
  y(0)=0,z^{'}(0)=0,
  z(0)=0,t^{'}(0)=1,t(0)=0$ and f given by 3.22.}}  
  \end{figure}
  
  
 
 \begin{figure}
 \epsfxsize= 3in
 \centerline{ \epsffile{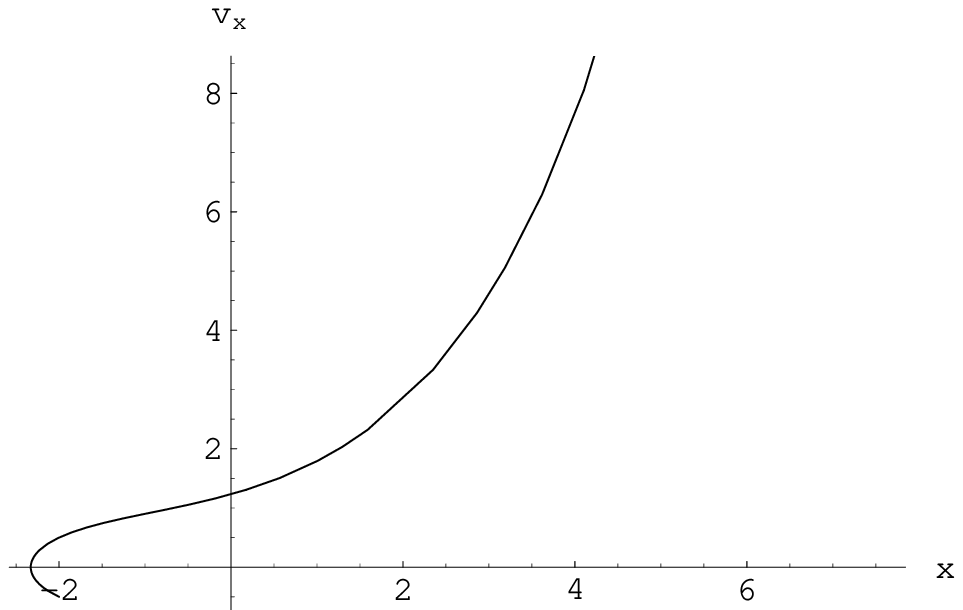}}
  
  \centerline{\caption{Phase space plot of a turning point solution 
  for a unit mass particle with initial conditions 
  $x(0)=2,x^{'}(0)=-0.5,y^{'}(0)=0.9188,y(0)=0,z^{'}(0)=0, 
  z(0)=0,t^{'}(0)=1,t(0)=0$.  On the left hand side of the domain wall 
  gravity is attractive, while on the right hand side of the domain 
  wall gravity is repulsive.  The space-time has parameters 
  $m=1,L=1/2, \mbox{and } k=1/2$ and metric given by $g$ in (4.15 and 
  4.25-4.27)}}
  \end{figure}
  
 \begin{figure}
 \epsfxsize= 3in
 \centerline{ \epsffile{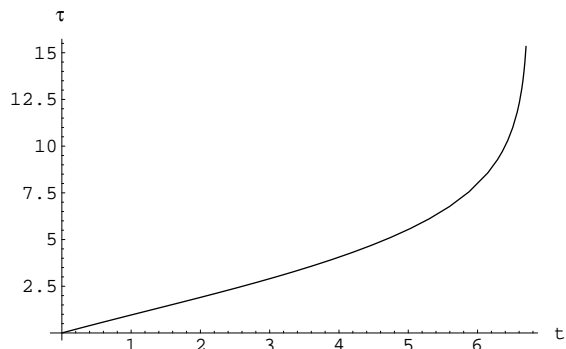}}
  
  \centerline{\caption{The red shift of $\tau(t)$ along a gedodesic 
  for a unit mass particle with initial conditions 
  $x(0)=2,x^{'}(0)=-0.5,y^{'}(0)=0.9188,y(0)=0,z^{'}(0)=0, 
  z(0)=0,t^{'}(0)=1,t(0)=0$.  The space-time has parameters 
  $m=1,L=1/2, \mbox{and } k=1/2$ and metric given by $g$ in (4.15 and 
  4.25-4.27 )}}
  \end{figure}
\end{document}